\documentclass[12pt]{iopart}

\begin{document}

\title{Do we understand the unquenched value of $f_B$?}

\author{Craig McNeile}

\address{Theoretical Physics Division, Dept. of Mathematical Sciences, 
          University of Liverpool, Liverpool L69 3BX, UK}
\ead{mcneile@amtp.liv.ac.uk}

\begin{abstract}
I review our qualitative understanding 
of the increase in the value of the B meson
decay constant ($f_B$), when dynamical fermions
are included in lattice QCD calculations.
\end{abstract}

\section{Introduction}

The over determination of the CKM matrix, using the new data from the
B factories at SLAC and KEK, is a sensitive test of the standard
model~\cite{Nir:2000py}. The determination of the CKM matrix elements
from experiment depends critically on hadronic parameters, such as
$f_B$, $B_B$, $B_{B_s}$ and $B_K$, most of which are inaccessible to
experiment, but can be calculated from lattice QCD.

In a recent survey of the latest lattice QCD results for the $f_B$
(the decay constant of the B meson) Bernard~\cite{Bernard:2000ki}
quotes: $f_B^{quenched} = 175(20)$ MeV from quenched QCD and $f_B =
200(30)$ MeV for the value of the B decay constant in full QCD.
Although all lattice QCD calculations~\cite{Bernard:2000ki} have seen
an increase in $f_B$ between quenched and unquenched QCD, the effect
in the world ``average'' is only at the one $\sigma$ level.  The
increase in $f_B$ between quenched and full QCD is more significant
for an individual collaboration's results, for example CP-PACS,
obtain~\cite{AliKhan:2000eg} $\frac{f_B^{n_f=2}}{f_B^{n_f=0}} =
1.11(6)$.  As unquenched simulations are so computationally demanding,
it seems useful to review the additional arguments that support the
increase in decay constants due to the inclusion of dynamical
fermions.

\section{What are unquenching effects?}

Lattice QCD is a ``clever'' finite difference
approximation to continuum QCD~\cite{Gupta:1999mn}.
Lattice QCD calculations involve computing the partition function
\begin{equation}
{\cal Z} =  \int \prod_{x} dU(x) 
\exp( -S_G ) (\det(M))^{n_f}
\label{eq:basicZdeat}
\end{equation}
where $U$ describe the gauge fields, $S_G$ is the 
lattice representation of the gauge action ($\frac{1}{4} F_{\mu\nu} F^{\mu\nu}$) and 
$M$ is a lattice representation of the Dirac 
operator for quarks. The quark fields have been
integrated out.
The dynamics of the gluon fields depends on the determinant of the
Dirac operator. The determinant describes the dynamics of the sea
quarks and is very expensive to compute.  Until recently, most
phenomenological lattice calculations did not include the determinant
in the dynamics (quenched QCD).  Surprisingly, quenched QCD
calculations describe experiment reasonably well.  The biggest recent
study of quenched QCD by the CP-PACS collaboration~\cite{Aoki:1999yr}
found that the spectrum of light hadrons disagreed with experiment 
at the 10\% level.

The results from an individual calculation depend on
the lattice spacing $a$ and physical volume.  These errors may be removed
by repeating the calculation at different lattice spacings and 
volumes and then extrapolating the results to the continuum and
infinite volume limit. For example, in the recent CP-PACS
calculation~\cite{AliKhan:2000eg}
 they obtain $f_B$ = $287(7)$ MeV, $234(8)$ MeV,
and $208(10)$ MeV, at lattice spacings: 0.22 fm, 0.16 fm, and 0.11 fm
respectively, from a calculation of $n_f =2 $ QCD. 
CP-PACS prefer to quote $f_B = $ $208(10)(11)$ MeV as their continuum
result, however Bernard~\cite{Bernard:2000ki} 
prefers to extrapolate the CP-PACS
data to the continuum assuming a  quadratic  dependence on
the lattice spacing (there are good, but not totally rigorous
arguments for this type of extrapolation)
and obtains $f_B = 190(12)(26)$ MeV. This kind of ambiguity
in the final analysis of lattice results is the cause of the 
large systematic errors in the final results of lattice calculations.
The cost of lattice QCD calculations that include
dynamical fermions
goes (something) like~\cite{Sharpe:1998hh}
$\frac{1}{a^{6.5}}$ where $a$ is the lattice spacing,
so halving the lattice spacing is very computationally
expensive.

\section{How to understand unquenching} \label{se:model}

There is a simple model of the effect of unquenching that is based on
the quark model~\cite{El-Khadra:1992ir,Butler:1994zx,Bernard:2000gd}.
Consider the Richardson heavy quark potential~\cite{Richardson:1979bt}.
\begin{equation}
V(q) \sim \frac{4 \pi}{(11 -2 \frac{n_f}{3})} 
\frac{1}{ q^2 \ln ( 1 + q^2/\Lambda^2) }
\label{eq:richardPOT}
\end{equation}
where $n_f$ is the number of flavours. 
Equation~\ref{eq:richardPOT} or the potential 
extracted from a lattice calculation
is used in Schr\"{o}dinger's equation
to calculate the wave function of mesons.
The decay constant is computed using the 
Van Royen-Weisskopf formulae
\begin{equation}
f \propto \mid \psi(0) \mid
\label{eq:VRQfdecay}
\end{equation}
In position space the potential in equation~\ref{eq:richardPOT} at
small radial separations is deeper in the full theory ($n_f$=3)
relative to the quenched theory ($n_f$=0).  So the decay constant
(computed from equation~\ref{eq:VRQfdecay}) is higher in the full
theory than in quenched QCD. There is evidence from many lattice QCD
calculations of the heavy quark potential, that dynamical fermions
produce a similar effect to the $n_f$ dependence of the Richardson 
potential~\cite{Bali:2000gf}.

The MILC collaboration~\cite{Bernard:2000gd} 
have systematically studied 
this model in an unquenched simulation using
staggered fermions. From the graphs in the 
paper~\cite{Bernard:2000gd}, at $m_{PS}/m_{V} \sim 0.58$,
the effect of unquenching is 3\% for $f_{\pi}$
and 7\% for $f_B$. The MILC collaboration found that
unquenching was smaller on the ratio of decay constants 
that on individual decay constants in this model.
A key qualitative prediction of this model is
that the decay constant in unquenched QCD is 
greater than the quenched value. Also the unquenching effect
should be larger for $f_B$ than $f_{\pi}$.

Another way to understand the effect of unquenching is to use a
quenched Lagrangian~\cite{Sharpe:1992ft,Bernard:1992mk}.  The idea is
similar to Chiral Lagrangian's, where a Lagrangian is written in terms
of hadron fields with the same symmetries as QCD. A number of unknown
parameters enter the Chiral Lagrangian, that must be fixed from
experiment (or from lattice calculations). Once the parameters are
fixed, the Lagrangian can be used to make predictions. For the case of
quenched QCD, ghost fields are introduced to cancel the quark
determinants. The resulting theory has a different symmetry to QCD,
but the idea is basically the same as for continuum chiral
perturbation theory.

The formalism for heavy-light mesons was developed by
Booth~\cite{Booth:1995hx} and Sharpe and Zhang~\cite{Sharpe:1996qp}.
The quenching errors are estimated by comparing the chiral logs (some
of the loop effects) in quenched and full QCD. This is thought to be
an upper bound on the magnitude of the quenching
effects~\cite{Sharpe:1996qp}.  Unfortunately, the estimate of the
quenching errors involves 7 parameters that are hard to determine
accurately. The message from this analysis was that $f_B$ in full QCD
could be either greater than or less than the quenched value. Also,
it was possible that the ratio $f_{B_s} / f_B$ had large 
quenching errors (this has not been found in 
simulations~\cite{Bernard:2000ki}).

\section{Lattice results for known decay constants}

The results of the first systematic studies of decay constants in
quenched QCD have been reviewed by Sharpe~\cite{Sharpe:1998hh}.
Sharpe concluded that $\frac{f_{\pi}}{m_{\rho}}$ was lower
than experiment, after the continuum limit had been taken.  The
recent large CP-PACS collaboration~\cite{Aoki:1999yr} study of
quenched QCD found that in quenched QCD $f_{\pi}^{quenched}$
($f_{K}^{quenched}$) = $120.0 \pm 5.7$ ($138.8 \pm 4.4$) MeV, that are
smaller than experiment by $2 \sigma$ ($5 \sigma$).

After the continuum extrapolation, the light decay constants from
quenched QCD are lower than experiment, consistent with the quark model in
section~\ref{se:model}. The SESAM collaboration~\cite{Eicker:1998sy}
found at one lattice spacing that the pion decay constant was
approximately one sigma larger with dynamical fermions, than from an
equivalent quenched simulation. The opposite trend was seen in the
QCDSF/UKQCD~\cite{Horsley:2000pz} data. It will be difficult to
determine the effect of dynamical fermions on light decay constants, 
until a continuum
extrapolation is done.

Including heavy quarks (Charm and Bottom) 
in lattice QCD calculations requires the 
introduction of new techniques, such as
effective field theories for the heavy 
quarks~\cite{Bernard:2000ki}. As a test of the
new heavy quark methods, the $f_{D_s}$ decay constant
is computed and compared against experiment.
The value for $f_{D_s}$ quoted in the particle
data table~\cite{Groom:2000in} is
$280 \pm 19 \pm 28 \pm 34$ MeV.
In table~\ref{tab:fdsSUMMARY}, I have collected some results for
$f_{D_s}$ from lattice gauge theory calculations (all the various
errors have been added in quadrature).  All the lattice results for
quenched QCD are lower than the experimental value.  This is
consistent with the picture that unquenching raises the value of a
decay constant.  The latest results~\cite{Bernard:2000ki} from lattice
calculations
show a 3-8\% increase in $f_{D_s}$, when dynamical
fermions are included.

\begin{table}[tb]
  \begin{tabular}{c|c|c|c}
Group  & Comments &  $f_{D_s}$  MeV & $\sigma$ \\ \hline
Average (lattice)~\cite{Draper:1998ms} & 
Review in 1998 & $220(25)$ & 2.2 \\ \hline
JLQCD~\cite{Ishikawa:1999xu}  & Continuum limit & 224(20) &
2.8 \\
MILC~\cite{Bernard:1998xi}    & Continuum limit & 210(27) &
2.6  \\
Collins et al.~\cite{Collins:2000ix}    & $a=0.18\;fm$ & 223(54) & 
1.1 \\ 
UKQCD~\cite{Bowler:2000xw}    & $a=0.068\;fm$ & 241(30) & 
1.3 \\ 
  \end{tabular}
  \caption{
Summary of recent quenched lattice results for \protect{$f_{D_s}$}.
The $\sigma$ column is the number of lattice errorbars
below the central experimental value of 280 MeV.
}
\label{tab:fdsSUMMARY}
\end{table}

\section{Conclusions}

A consistent picture does seem to emerge from lattice QCD
calculations, that unquenching does raise the value of decay
constants. However, the errors on the calculations need
to be reduced. Ideally, a calculation of $f_B$ similar to the 
recent CP-PACS calculation is required~\cite{AliKhan:2000eg},
but at lighter sea masses and smaller lattice spacings.
Improvements in lattice techniques and faster computers
will reduce the errors on $f_B$.


\end{document}